\documentclass[12pt]{iopart}
\usepackage{graphicx}
\usepackage{color}
\begin{document}

\title[Relativistic coupled-cluster study of BaF]{Relativistic coupled-cluster study of BaF in search of $\mathcal{CP}$ violation}

\author{Kaushik Talukdar$^{1,}$\footnote{talukdar.kaushik7970@gmail.com}, Malaya K. Nayak$^{2,}$\footnote{mk.nayak72@gmail.com},
Nayana Vaval$^{3,}$\footnote{np.vaval@gmail.com}, and Sourav Pal$^{4,1,}$\footnote{spal@chem.iitb.ac.in}}
\address{$^1$ Department of Chemistry, Indian Institute of Technology Bombay, Powai, Mumbai 400076,  India}
\address{$^2$ Theoretical Chemistry Section, Bhabha Atomic Research Centre, Trombay, Mumbai 400085, India}
\address{$^3$ Electronic Structure Theory Group, Physical Chemistry Division, CSIR-National Chemical Laboratory, Pune 411008, India}
\address{$^4$ Indian Institute of Science Education and Research Kolkata, Mohanpur 741246, India}

\vspace{10pt}
\begin{indented}
\item[]November 2019
\end{indented}

\begin{abstract}
BaF is one of the potential candidates for the experimental search of the electric dipole moment of the electron (eEDM).
The NL-eEDM collaboration is building a new experimental set up to measure the eEDM using the BaF molecule [The NL-eEDM collaboration, Eur. Phys. J. D (2018) 72: 197].
To analyze the results of such an experiment, one would require the accurate value of the molecular ${\mathcal{P,T}}$-odd interaction parameters 
that cannot be measured from any experiment. In this work, we report the precise value of the ${\mathcal{P,T}}$-odd interaction parameters of the BaF molecule
obtained from the four-component relativistic coupled-cluster calculations. We also calculate the hyperfine structure (HFS) constants of the same molecule
to assess the reliability of the reported molecular parameters. The calculated HFS constants show good agreement with the available experimental values. Further,
the systematic effects of electron-correlation along with the roles of inner-core electrons and the virtual energy functions in the
calculation of the studied properties of BaF are investigated.

\end{abstract}

%
%
%
%
%

\maketitle
\section{Introduction}
BaF, being a heavy polar open-shell molecule, shows excellent potential as a candidate in search of the electric
dipole moment (EDM) of an electron. Recently, a new experiment in search of the electric dipole moment of the electron (eEDM) using the BaF
molecule is proposed by the NL-eEDM collaboration \cite{nl_eedm}. (Lifetimes of the molecular states relevant for the experimental search of
the eEDM in BaF are recently measured \cite{nl_eedm2}.)
Observation of the eEDM is a signature of violation of time-reversal ($\mathcal{T}$) as well as parity ($\mathcal{P}$) invariance. $\mathcal{T}$-violation
is equivalent to charge conjugation ($\mathcal{C}$) and parity violation according to the $\mathcal{CPT}$ theorem that states that the universe remains invariant under the combined operation of $\mathcal{CPT}$. (Thus we shall
often use the term ``$\mathcal{P,T}$-odd properties'' as a synonym for ``$\mathcal{CP}$-violating properties'' throughout this paper.)
The ${\mathcal{CP}}$ violation is one of the several 
conditions that can explain the matter-antimatter asymmetry of the universe \cite{sakharov}. However, ${\mathcal{CP}}$ or ${\mathcal{T}}$ violation
in the standard model (SM) of elementary particles miserably fails to justify the observed imbalance between matter and antimatter in the present-day universe. 
The best upper bound limit of eEDM ($< 1.1 \times 10^{-29}$ e.cm) to date is obtained from the ThO-experiment \cite{tho_new}. The current
experimental sensitivity on the eEDM cannot be explained from the ${\mathcal{CP}}$-violating mechanism within the SM. The eEDM is predicted to be 
almost zero ($\approx 10^{-38}$ e.cm) in the standard model, up to the three-loop level.
That is why new sources of symmetry violation beyond the SM, which is nowadays known as ``new physics'' is becoming the key interest to the physicists.
The eEDM \cite{bernreuther_1991, tl_edm, ybf_edm, titov_tho, tho_edm} along with the other $\mathcal{P,T}$-odd violating properties such as the 
EDM of nucleon \cite { pospelov_2005, engel_2013}, the scalar-pseudoscalar (S-PS) nucleon-electron neutral current coupling \cite{tho_edm, sasmal_raf, kudashov_2014, sudip_hgh},
the nuclear magnetic quadrupole moment (MQM) \cite{fleig_tan, flambaum_2017, titov_2014} etc., can contribute to the permanent EDM of an atom or a molecule.
In open-shell diatomics, the eEDM and the S-PS nucleon-electron interaction are the major sources of permanent EDM. The open-shell
molecules containing heavy deformed nuclei of spin $I > 1/2$ can also produce additional permanent EDM due to the nuclear MQM effects. 
However, the EDM due to the MQM effects in molecules with spherical or nearly spherical nuclei is usually insignificant. 
In the present work, we limit our study only to the eEDM and the S-PS nucleus-electron coupling interactions. 

A nonzero EDM of an electron exists if and only if it has an aspherical charge distribution along the spin axis. This is only possible
when there is a violation of $\mathcal{T}$ invariance along with $\mathcal{P}$. As mentioned above the eEDM predicted from the SM is much
smaller than the current experimental limit, a measurable eEDM is only possible from the $\mathcal{P,T}$-odd mechanism beyond the SM.
The predicted value of the eEDM in many new models lies in the current experimental sensitivity. 
The S-PS nucleus-electron interaction, on the other hand, arises
as a result of coupling between the scalar hadronic current and the pseudoscalar electronic current, which can be mediated via scalar
and pseudoscalar components of a neutral Higgs Boson particle \cite{barr_1992}.
As the known Higgs Boson in the SM is a scalar, it can not explain such interaction.
However, numerous multi-Higgs models (along with the minimal supersymmetric standard model) can predict this type of interaction.
These models can also predict the nonconservation of baryon number which is another condition for the large imbalance of matter and antimatter
in our universe \cite{kazarian_1992}. That is why the measurement of the eEDM and the S-PS nucleon-electron interaction is very significant. Due to the presence of
large effective electric field ($E_\mathrm{eff}$) in the heavy polar open-shell diatomic molecular systems, they are suitable candidates for
the low energy eEDM experiments. In such an experiment, the energy shift caused by the interaction of the permanent molecular EDM with
the applied electric field is measured. However, to interpret the experimental results in terms of the eEDM and the fundamental S-PS nucleus-electron coupling
constant, one needs the accurate value of the molecular parameter associated with the respective $\mathcal{P,T}$-odd mechanism.
But no experiment can provide the knowledge of those parameters. However, the molecular parameters can be calculated by {\it ab initio}
methods. The $\mathcal{P,T}$-odd interaction parameters are very much dependent on the electron density at the vicinity of the heavy nucleus of
the diatomic molecule. The same is also true for the hyperfine structure (HFS) interaction constants. This means that properties like the $\mathcal{P,T}$-odd
molecular parameters and the HFS coupling constants are sensitive to the wave function near the heavy nucleus of the molecule. Thus, these properties are also
known as the core properties or sometimes as the ``atom-in-compound'' (AIC) properties \cite{titove_aic}. For the accurate calculation of such a property, the effects of the relativistic
motion of electrons and the electron-correlation are extremely important. Therefore, an {\it ab initio} method that can efficiently incorporate both 
the relativistic and the correlation effects of electrons should be suitable for such a calculation. It is worth mentioning that the accurate calculation of an AIC property
is often very difficult because of the strong inter-electronic correlation effects present in the molecules. So, one needs to understand the role of electron-correlation 
in such molecular calculations to obtain the accurate values of the core properties.

The main objective of this article is to precisely calculate the ${\mathcal{P}}$,${\mathcal{T}}$-odd interaction constants
{\it viz.}, $E_\mathrm{eff}$ and $W_\mathrm{s}$ of BaF in its ground electronic ($^2\Sigma_{\frac{1}{2}}$) state using the $Z$-vector method within 
the four-component relativistic coupled-cluster singles and doubles (CCSD) model \cite{sasmal_pra_rapid}. 
Although in recent times, the ${\mathcal{P}}$,${\mathcal{T}}$-odd constants
of this molecule have been reported from semi-empirical \cite{ kozlov_1995} and various {\it ab initio} 
calculations \cite{sunaga_HgF, kozlov_baf, nayak_2006, nayak_2007, nayak_2008, berger_2017, meyer_2006, fukuda_2016}, the calculation of the same using a
higher level of theory is always important. 
The accuracy of the $\mathcal{P,T}$-odd molecular parameters can not be directly 
assessed since they can not be experimentally measured. However, there is an indirect way to do so by
calculating the HFS constants and then comparing them with the available experimental values. 
This is because the accuracy in the calculated HFS constants, similar to the $\mathcal{P,T}$-odd molecular parameters, depends on the accuracy of 
the wavefunction near the nuclear region.
Thus, we report the hyperfine structure constants of the BaF molecule to assess the reliability of the calculated ${\mathcal{P}}$,${\mathcal{T}}$-odd 
interaction constants. Further, the systematic effects of electron-correlation as well as the roles of inner-core electrons 
and the virtual spinors in the molecular calculations of BaF are discussed in this article.

\par 

The paper is organized as follows: The theory of the calculated properties and
the relativistic coupled-cluster method is discussed in Sec. \ref{theory}.
In Sec. \ref{comp}, computational details are given. The Sec. \ref{res_dis} deals with the discussion of the results.
Finally, the concluding remark is given in Sec. \ref{conc}. Atomic units are used here unless stated. 

\section{Theory}\label{theory}
The Hamiltonian for the interaction of eEDM ($d_e$) with the internal molecular electric field \cite{kozlov_1987, titov_2006} is
$H_d = 2icd_e \gamma^0 \gamma^5 {\bf \it p}^2$,
where $c$ is the speed of light, $\gamma$ are the usual Dirac matrices and {\bf \it p} is the momentum operator. Now, the $E_\mathrm{eff}$ can be defined as
\begin{eqnarray}
 E_{\mathrm{eff}} = | \langle \Psi_{\Omega} | \sum_j^n \frac{H_d(j)}{d_e} | \Psi_{\Omega} \rangle |.
 \label{E_eff}
\end{eqnarray}
Here $\Psi_{\Omega}$ is the wavefunction of the state $\Omega=1/2$, which is the projection of total electron angular momentum of BaF on the molecular axis (z-axis) and $n$ is 
the total number of electrons.

\par
The interaction Hamiltonian for S-PS nucleus-electron coupling \cite{hunter_1991} is defined as
$H_{\mathrm{SP}}= i(G_{F}/\sqrt{2})Zk_{s} \gamma^0 \gamma^5 \rho_N(r)$,
where G$_F$ is the Fermi constant, $Z$ is the number of protons, and $\rho_N(r)$ is the nuclear charge density normalized to unity. The fundamental S-PS coupling constant
$k_s$ is defined as Z$k_s$=(Z$k_{s,p}$+N$k_{s,n}$), where N is the number of neutrons, and $k_{s,p}$ and $k_{s,n}$
are the electron-proton and electron-neutron coupling constant, respectively. The parameter
$W_{\mathrm{s}}$ can be evaluated by
\begin{eqnarray}
 W_{\mathrm{s}}=|\frac{1}{\Omega k_\mathrm{s}}\langle \Psi_{\Omega}|\sum_j^n H_{\mathrm{SP}}(j)| \Psi_{\Omega} \rangle|.
\label{W_s}
\end{eqnarray}

\par


The parallel ($A_{\|}$) and perpendicular ($A_{\perp}$) components of the HFS constant
of a diatomic molecule can be defined as
\begin{eqnarray}
A_{\|(\perp)}= \frac{\vec{\mu_k}}{I\Omega} \cdot \langle \Psi_{\Omega} | \sum_i^n
\left( \frac{\vec{\alpha}_i \times \vec{r}_i}{r_i^3} \right)_{z(x/y)} | \Psi_{\Omega(-\Omega)}  \rangle,
\label{hfs_mol}
\end{eqnarray}
where $\vec{\mu}_k$ is the magnetic moment of the nucleus $k$ and $\vec{\alpha}_i$ is Dirac matrix for $i$th electron.
\par

To calculate the above mentioned properties, we use the relativistic coupled-cluster singles and doubles (CCSD) wavefunction, 
which is given by $|\Psi_{cc}\rangle=e^{T}|\Phi_0\rangle$. Here $\Phi_0$ is the four-component Dirac-Hartree-Fock wavefunction and
$T=T_1+T_2$, is known as excitation operator which is defined as
\begin{eqnarray}
 T_m= \frac{1}{(m!)^2} \sum_{ij\dots ab \dots} t_{ij \dots}^{ab \dots}{a_a^{\dagger}a_b^{\dagger} \dots a_j a_i} ,
\end{eqnarray}
where $i,j(a,b)$ indices refer to the occupied (unoccupied) spinors, and $a_p^{\dagger}$ ($a_p$) refers to the creation (annihilation) operator for
spinor $p$. $t_{ij..}^{ab..}$ is the cluster amplitude corresponding 
to $T_m$.
The unknown amplitudes corresponding to $T_1$ and $T_2$ are solved 
using the following equations:
\begin{eqnarray}
 \langle \Phi_{i}^{a} | (H_Ne^T)_c | \Phi_0 \rangle = 0 , \,\,
  \langle \Phi_{ij}^{ab} | (H_Ne^T)_c | \Phi_0 \rangle = 0 ,
 \label{cc_amplitudes}
\end{eqnarray}
where $H_N$ is the normal ordered Dirac-Coulomb Hamiltonian, and $\Phi_{i}^{a}$ and $\Phi_{ij}^{ab}$ are
the single and double excited determinants with reference to $\Phi_0$, respectively.
The subscript $c$ means that only the connected terms survive in the contraction between $H_N$ and $T$. 

\par

\par
Recently, Sasmal {\it et al.} \cite{sasmal_pra_rapid} extended the $Z$-vector technique into 
the four-component relativistic CCSD framework and successfully implemented the method to precisely calculate various AIC properties of molecules
\cite{sasmal_raf, sudip_hgh, sudip_pbf, sudip_pbf_pt}.
For more detailed knowledge of the $Z$-vector method, one can see Refs. \cite{schafer_1984, zvector_1989, lambda_1990}.
The perturbation independent linear operator used in the $Z$-vector method within the CCSD framework is
$\Lambda=\Lambda_1+\Lambda_2$,
where
\begin{eqnarray}
 \Lambda_m= \frac{1}{(m!)^2} \sum_{ij \dots ab \dots} \lambda_{ab \dots}^{ij \dots}{a_i^{\dagger}a_j^{\dagger} \dots a_b a_a} ,
\end{eqnarray}
where $\lambda_{ab \dots}^{ij \dots}$ is the amplitude corresponding to $\Lambda_m$.
The explicit equations for the amplitudes of $\Lambda_1$
and $\Lambda_2$ are
\begin{eqnarray}
\langle \Phi_0 |[\Lambda (H_Ne^T)_c]_c | \Phi_{i}^{a} \rangle + \langle \Phi_0 | (H_Ne^T)_c | \Phi_{i}^{a} \rangle = 0,
\end{eqnarray}
\begin{eqnarray}
\langle \Phi_0 |[\Lambda (H_Ne^T)_c]_c | \Phi_{ij}^{ab} \rangle + \langle \Phi_0 | (H_Ne^T)_c | \Phi_{ij}^{ab} \rangle \nonumber \\
 + \langle \Phi_0 | (H_Ne^T)_c | \Phi_{i}^{a} \rangle \langle \Phi_{i}^{a} | \Lambda | \Phi_{ij}^{ab} \rangle = 0.
\label{lambda_2}
\end{eqnarray}
Once the amplitudes are known, the energy derivative (or the desired property) can be obtained by
\begin{eqnarray}
 \Delta E' = \langle \Phi_0 | (O_Ne^T)_c | \Phi_0 \rangle + \langle \Phi_0 | [\Lambda (O_Ne^T)_c]_c | \Phi_0 \rangle,
\end{eqnarray}
where, $O_N$ is known as the derivative of normal ordered perturbed Hamiltonian with respect to external field of perturbation.

\section{Computational details}\label{comp}
We perform our calculations using a locally modified version of DIRAC17 \cite{dirac17}, which is interfaced with the $Z$-vector code developed in our group.
The default parameters for the nuclei in DIRAC are used as the nuclear parameters in our calculations where the nuclei are modeled by the Gaussian charge distribution \cite{visscher_1997}. 
The small basis sets are generated from the large basis using the restricted kinetic balance \cite{dyall_2007} condition. 
In our calculation, the negative energy spectrum is projected out using the ``no virtual pair approximation'' \cite{nvpa_2016}. 
The bond length of BaF is taken as 2.16 \AA \cite{srf}. We use the following uncontracted Gaussian 
basis sets: double-zeta (DZ): dyall.cv2z \cite{dyall_2004} for Ba, cc-pCVDZ \cite{dunning_1989} for F, 
triple-zeta (TZ) basis: dyall.cv3z \cite{dyall_2004} for Ba and cc-pCVTZ \cite{dunning_1989} for F; quadruple-zeta (QZ) basis: dyall.cv4z \cite{dyall_2004} for Ba 
and cc-pCVQZ \cite{dunning_1989} for F. 
We correlate all the electrons in most of the calculations, however, 30 inner-core (1$s$-3$d$)electrons are excluded from correlation treatment for the frozen-core calculations (see Table 4).
We cut off the virtual spinors having energy more than a certain value. 
The information about the basis sets and the cutoffs used for virtual spinors are given in Table \ref{basis}. 

\begin{table}[ht]
\caption{ Cutoffs for virtual spinors and basis sets used in our calculations.}
\newcommand{\mc}[2]{\multicolumn{#1}{#2}}
\begin{center}
\begin{tabular}{lccccr}
\hline 
\mc{4}{c}{Basis} & \mc{2}{c}{Virtual}\\
\cline{1-4} \cline{5-6}  
Name & Nature & Ba & F & Cutoff (a.u.) & Spinors \\
\hline
A & DZ & dyall.cv2z & cc-pCVDZ & 500 & 201 \\      
B & TZ &  dyall.cv3z & cc-pCVTZ & 200 & 325 \\
C & TZ & dyall.cv3z & cc-pCVTZ & 500 & 361 \\
D & QZ & dyall.cv4z & cc-pCVQZ & 200 & 579 \\   
\hline
\end{tabular}
\end{center}
\label{basis}
\end{table}

\section{Results and discussion}\label{res_dis}
\begin{figure*}[ht]
\centering
\begin{center}
\includegraphics[scale=.7]{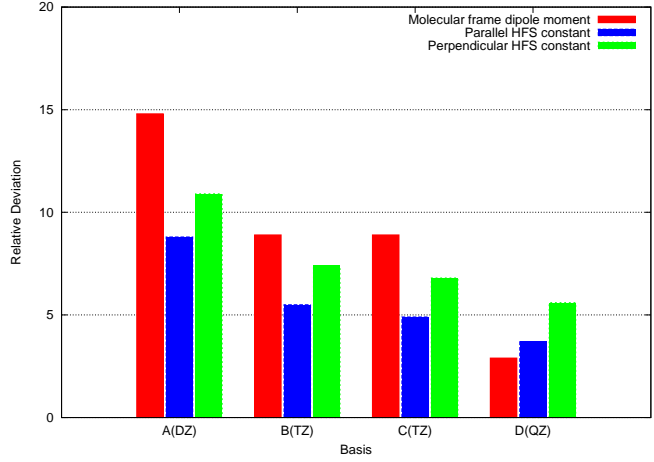}
\vspace{1cm}
\caption {Relative deviation of the calculated molecular frame dipole moment ($\mu$), parallel HFS constant ($A_{\|}$) and perpendicular HFS constant ($A_{\perp}$) from the corresponding experimental values for BaF with different basis sets}
\label{baf_dev}
\end{center}  
\end{figure*}

The molecular frame dipole moment ($\mu$), and the parallel and perpendicular components of the HFS constant
of $^{137}$Ba in BaF calculated with different basis sets are presented in Table \ref{baf_hfs}. Our results are compared with the
available experimental values \cite{baf_dipole, baf_HFS} as well as other theoretically calculated values \cite{kozlov_baf, nayak_2006, fukuda_2016}
in the same table. The magnitude of both the dipole moment and the HFS constants increases with the use of a higher quality basis set. 
This trend is expected as the addition of the higher angular momentum basis functions and the high-energy virtual 
functions usually improve the correlation space, which in turn, provide some improvement in the calculated results. However, the use of basis
C over B does not improve the molecular dipole moment of BaF which is probably due to the fact that both B (TZ, 200 a.u.) and C (TZ, 500 a.u.) basis sets are of triple-zeta quality and 
the additional high-lying virtual energy functions present in basis C contribute negligibly to the molecular frame dipole moment. The relative 
deviation ($ = |\frac{Theory-Experiment}{Theory}| \times 100\%$) of the calculated properties from the experimental values are 
shown in Fig. \ref{baf_dev}. From Table \ref{baf_hfs} and Fig. \ref{baf_dev}, we observe that the values obtained with basis D (QZ, 200 a.u.) show the least
relative deviation. Our HFS constants using this basis set are in better agreement with the experimental results than those reported by Kozlov {\it et al.} \cite{kozlov_baf}
using the generalized relativistic effective core potential (GRECP) method with effective operator (EO) based perturbative corrections 
within the restricted active space self-consistent field (RASSCF) framework and by Fukuda {\it et al.} \cite{fukuda_2016} employing the restricted
active space configuration interaction (RASCI) singles and doubles method. On the other hand, the molecular frame dipole moments calculated by
Fukuda {\it et al.} (including all the electrons in the RASCI active space) \cite{fukuda_2016} and Nayak {\it et al.} (including only 17 electrons in the active space) \cite{nayak_2006} 
using the RASCI method show better agreement with the experiment than that reported by us.

\begin{table*}[ht]
\caption{Molecular frame dipole moment  $\mu$ (in Debye), and magnetic HFS constants (in MHz) of BaF }
\newcommand{\mc}[1]{\multicolumn{#1}}
\begin{center}
\begin{tabular}{lccr}
\hline
Basis/Method & $\mu$ & \mc{2}{c}{$^{137}$Ba} \\
\cline{3-4}  
 &  & A$_{\|}$ & A$_{\perp}$ \\
\hline
A & 2.76 & 2254 & 2165 \\
B &  2.91 & 2326 & 2236 \\
C & 2.91 & 2338 & 2247 \\
D & 3.08 & 2365 & 2274 \\
\hline
GRECP/RASSCF/EO \cite{kozlov_baf} & - & 2272 & 2200 \\
RASCI(17e) \cite{nayak_2006} & 3.20  & - & - \\
RASCI(65e) \cite{fukuda_2016} & 3.11 & 1589 & - \\
\hline
Experiment & 3.17(3) \cite{baf_dipole} & 2453(9) \cite{baf_HFS} & 2401(6) \cite{baf_HFS}\\
\hline
\end{tabular}
\end{center}
\label{baf_hfs}
\end{table*}
\begin{figure*}[ht]
\centering
\begin{center}
\includegraphics[scale=.9]{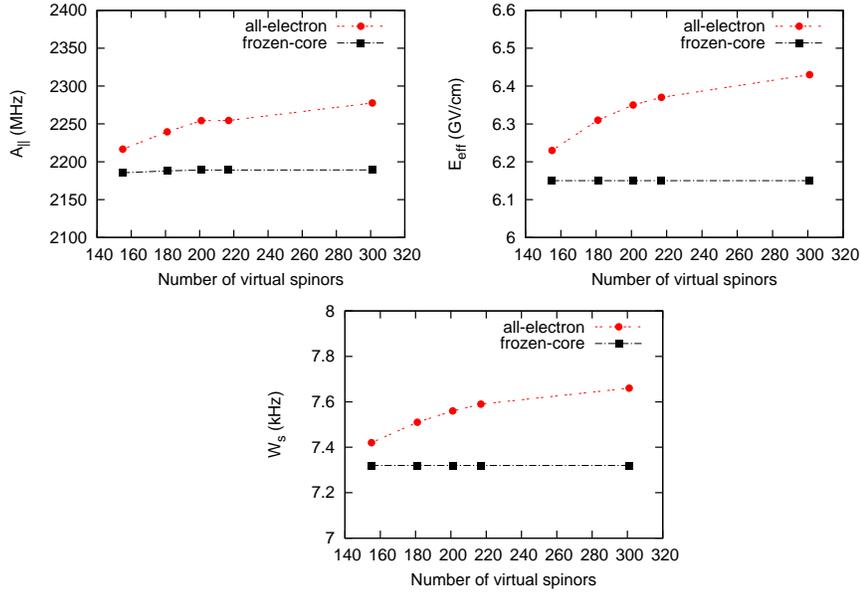}
\vspace{1cm}
\caption {Convergence of the AIC properties of BaF with the number of virtual spinors}
\label{Ws_convergence}
\end{center}  
\end{figure*}
\begin{table}[ht]
\caption{${\mathcal{P,T}}$-odd interaction constants (E$_\mathrm{eff}$ in GV/cm, W$_\mathrm{s}$ in kHz, and R in 10$^{18}$/e.cm unit) of BaF}
\begin{center}
\begin{tabular}{lcccr}
\hline
Basis & Nature & $E_\mathrm{eff}$ & $W_\mathrm{s}$ & R=$E_\mathrm{eff}$/W$_\mathrm{s}$ \\ 
\hline
A & DZ & 6.35 & 7.56 & 203.1 \\
B & TZ & 6.56 & 8.30 & 191.1 \\
C & TZ & 6.59 & 8.34 & 191.1 \\
D & QZ & 6.52 & 8.35 & 188.8 \\
\hline
\end{tabular}
\end{center}
\label{baf_pt}
\end{table}
In Table \ref{baf_pt}, we present the $\mathcal{P,T}$-odd interaction constants $E_\mathrm{eff}$ and $W_\mathrm{s}$ along with their 
ratio ($R$) for the BaF molecule. The effective electric field in BaF increases from 6.35 to 6.56 and then to 6.59 GV/cm when we go from basis A (DZ, 500 a.u.)
to basis B (TZ, 200 a.u.) and then to basis C (TZ, 500 a.u.). On the contrary, as we go from basis C to basis D (QZ, 200 a.u.), $E_\mathrm{eff}$ decreases from 6.59 to 6.52
GV/cm. The S-PS nucleon-electron interaction parameter increases from 7.56 to 8.30 and then to 8.34 kHz while going from basis A to basis B and then to basis C.
However, a very small (0.01 kHz) increment in $W_s$ is observed for the use of basis D over C. On the other hand, the ratio ($R$) of $E_\mathrm{eff}$ to $W_\mathrm{s}$ decreases
from 203.1 to 188.8 in the unit of 10$^{18}$/e.cm with the use of basis A to D.
It is noteworthy that the virtual functions with energy in-between 200 a.u. and 500 a.u. are absent in basis D unlike basis A and C.
We want to reiterate that the use of TZ basis (both B \& C) over DZ basis (A)
results in an increase in the magnitude of the studied AIC properties, but the use of basis D (QZ, 200 a.u.) over the TZ basis sets does not reflect 
the similar trend, especially in the case of $E_\mathrm{eff}$. Unlike in the cases of the HFS constants and $W_s$, 
the high angular momentum basis functions probably decrease the magnitude of the effective electric field of BaF, 
especially when we go from TZ to QZ basis. A similar observation was also inferred in Ref. \cite{talukdar_hgf} for the HgF molecule.
In such a case, the addition of high-lying virtual spinors into the configuration space may compensate for the effects of the high angular 
momentum basis functions. 
Nevertheless, as the accuracy of the HFS results reflects the reliability of the $\mathcal{P,T}$-odd interaction parameters and the 
relative deviation of the HFS constant calculated using basis D is the least, we consider the property values calculated using this basis set 
to be our most reliable results. Thus, our most reliable results for $E_\mathrm{eff}$, $W_\mathrm{s}$ and $R$ in BaF are 6.52 GV/cm, 8.35 kHz and 
188.8$\times$10$^{18}$/e.cm, respectively. One should note, since we have cut off the virtual spinors of energy more than 200 a.u. for basis D, 
it is important to estimate the contribution of the missing high-energy virtual spinors to the calculated properties to achieve more precision. 
(The contribution of the virtual spinors to the core properties of BaF is discussed at the end of this section.)

The convergence of the AIC properties of BaF with the number of virtual spinors is shown in Table \ref{core_effect}. Two sets of calculations
are performed to understand the effect of virtual spinors and that of core-correlation in the study of AIC properties of BaF.
Firstly, we correlate all the electrons in the molecular calculation to report $A_{\|}$, $E_\mathrm{eff}$ and $W_s$ at different cutoffs for virtual
spinors using the DZ (dyall.cv2z for Ba and cc-pCVDZ for F) basis set. Secondly, the same calculations are repeated by correlating only 35 
outer-electrons (i.e., freezing the 30 inner-core (1$s$-3$d$) electrons) to see the correlation trend of the inner-core electrons.
The role of core-correlation effects can be understood by comparing the property-value obtained from the all-electron correlation calculation
and the frozen-core molecular calculation.
We see from Table \ref{core_effect} that the magnitude of AIC
properties gradually increases with the increase in the number of virtual energy functions when all the electrons of BaF are correlated.
Unlike the all-electron correlation calculation, the magnitude of the properties obtained from the frozen-core calculation increases negligibly with the number of virtual spinors. 
This means that the trend of the AIC properties of BaF with the number of virtual spinors in a given basis set is not similar for the all-electron and the frozen-core correlation 
calculations. As observed in Table \ref{core_effect}, the inclusion of
the high-energy virtual functions in molecular calculation hardly affects the property-value for the frozen-core case. In other words, the high-energy virtual
functions are important only for the proper correlation of the inner-core electrons. This inference is consistent with the findings of Refs. \cite{core_jcp, core_pra,talukdar_hfs}.
Furthermore, the correlation of the 30 inner-core (1$s$-3$d$) electrons contributes 3.88\% to $A_{\|}$, 4.35\% to the effective electric field
and 4.44\% to the S-PS interaction parameter when all the virtual spinors are included in the molecular calculation of BaF. Thus, the correlation of 
the inner-core electrons along with the high-lying virtual spinors play crucial roles in the precise calculations of the AIC properties.

\begin{table}[ht]
\caption{The AIC properties of BaF at various cutoffs for virtual spinors. (Basis used: dyall.cv2z for Ba and cc-pCVDZ for F.) }
\newcommand{\mc}[1]{\multicolumn{#1}}
\begin{center}
\begin{tabular}{lccccr}
\hline
\mc{2}{c}{Virtual} & Occupied & A$_{\|}$ & $E_\mathrm{eff}$ & $W_\mathrm{s}$ \\
\cline{1-2} 
Cutoff(a.u.) & Spinors & Spinors & (MHz) & (GV/cm) & (kHz) \\
\hline
50 & 155 & 65  & 2216.6 & 6.23 & 7.42 \\
200 & 181 & 65  & 2239.5 & 6.31 & 7.51 \\
500 & 201 & 65  & 2254.3 & 6.35 & 7.56 \\
1000 & 217 & 65  & 2254.5 & 6.37 & 7.59 \\
No cutoff & 301 & 65 & 2277.6 & 6.43 & 7.66 \\
No cutoff (including & 301 & 65 & 2278.3 & 6.33 & 7.68 \\
Gaunt interaction) & & & & & \\
\hline 
50 & 155 & 35 & 2185.4 & 6.15 & 7.32 \\
200 & 181 & 35 & 2187.9 & 6.15 & 7.32 \\
500 & 201 & 35 & 2189.0 & 6.15 & 7.32 \\
1000 & 217 & 35 & 2189.0 & 6.15 & 7.32 \\
No cutoff & 301 & 35  & 2189.2 & 6.15 & 7.32 \\
\hline
\end{tabular}
\end{center}
\label{core_effect}
\end{table}
\begin{table}[ht]
\caption{Comparison of ${\mathcal{P,T}}$-odd interaction constants (E$_\mathrm{eff}$ in GV/cm and W$_\mathrm{s}$ in kHz) of BaF calculated by various methods.} 
\begin{center}
\begin{tabular}{lccr}
\hline
Method  & $E_\mathrm{eff}$ & $W_\mathrm{s}$ \\ 
\hline
$Z$-vector CCSD (this work) & 6.52 & 8.35 \\ 
LECCSD \cite{sunaga_HgF} & 6.6 & 8.4 \\ 
GRECP/RASSCF \cite{kozlov_baf} & 4.63 & 5.9 \\
GRECP/RASSCF/EO \cite{kozlov_baf} & 7.53 & -  \\
RASCI (17e) \cite{nayak_2006, nayak_2007} & 7.28 & 9.7  \\
RASCI (65e) \cite{fukuda_2016} & 4.56 &  \\
MBPT \cite{nayak_2008} & - & 8.4  \\
GHF-ZORA \cite{berger_2017} & 6.82 & - \\
GKS-ZORA \cite{berger_2017} & 5.99 &  - \\
NR-MRCI \cite{meyer_2006} & 5.17 & - \\
Semi-empirical \cite{kozlov_1995} & 7.24 & 11.0 \\
\hline
\end{tabular}
\end{center}
\label{comp_pt}
\end{table}
In Table \ref{comp_pt}, we compare the $\mathcal{P,T}$-odd molecular parameters of BaF reported in this work with the literature 
values \cite{sunaga_HgF, kozlov_baf, nayak_2006, nayak_2007, nayak_2008, berger_2017, meyer_2006, kozlov_1995, fukuda_2016}.
Das and coworkers \cite{sunaga_HgF} calculated the $\mathcal{P,T}$-odd constants of BaF using relativistic linear expectation-value coupled cluster (LECCSD) method. The
LECCSD method lacks the non-linear terms in the expectation-value expression \cite{sunaga_HgF} and misses some of the correlation effects. 
Their reported values of $E_\mathrm{eff}$ and $W_s$ are in good agreement with our results though they 
used the triple zeta basis set with 80 a.u. as the cutoff for virtual spinors in their calculations.
Kozlov {\it et al.} \cite{kozlov_baf} performed the first {\it ab initio} calculation for the $\mathcal{P,T}$-odd constants of BaF employing the GRECP/RASSCF/EO method.
Nayak {\it et al.} \cite{nayak_2006, nayak_2007} reported $E_\mathrm{eff}$ as well as $W_s$ using the restricted active space
configuration interaction (RASCI) method (including only 17 electrons in the active space) and Fukuda {\it et al.} \cite{fukuda_2016} calculated $E_\mathrm{eff}$ employing 
the RASCI method (including 65 active electrons with DZ basis). Nayak {\it et al.} \cite{nayak_2008} also performed relativistic second-order many-body
perturbation theory (MBPT) computation in BaF to report the S-PS interaction constant, which shows good agreement with our value.
Recently, Gaul and Berger \cite{berger_2017} calculated the $\mathcal{P,T}$-odd parameter related to the eEDM interaction in the BaF molecule employing a quasi-relativistic 
two-component approach namely, a complex generalized Hartree-Fock (GHF) and a complex generalized Kohn-Sham (GKS) scheme within the zero-order regular approximation (ZORA).
The non-relativistic multi-reference configuration interaction (NR-MRCI) result \cite{meyer_2006} for the effective electric field of BaF is smaller than our value.
On the other hand, the semi-empirical values of the $\mathcal{P,T}$-odd parameters in BaF 
reported by Kozlov {\it et al.} \cite{kozlov_1995} are larger than those reported in our work. 
Nevertheless, our results are more reliable than the other theoretical values because we employ energy-derivative approach within the four-component
relativistic CCSD framework with a sufficiently high-quality relativistic basis sets in the molecular calculation. The good agreement of the HFS constants
with the experimental results (see Table \ref{baf_hfs}) reflects this fact.

As mentioned above, our most reliable results for BaF are calculated at its equilibrium bond length with basis D (i.e., QZ, 200 a.u. as a cutoff for virtual spinors)
using the DC Hamiltonian in four-component relativistic CCSD method. We thus neglect the role of high-lying virtual spinors, the vibrational effects, 
the higher-order correlation contribution and the Breit/QED effects, etc. in our molecular calculations. As one would need the values of $\mathcal{P,T}$-odd 
molecular parameters of BaF as accurate as possible to analyze the experimental findings of the eEDM experiment, the inclusion of the said effects in the molecular
calculations could be important. However, in this work, it is not possible to perform molecular calculations with QZ basis sets incorporating all those effects. 
Instead, we estimate the possible errors with a lower quality (DZ) basis set. We can estimate the error due to restriction of 
the correlation space (since we cut off the virtual spinors with energy above 200 a.u. in basis D) from Table \ref{core_effect}.
Inclusion of the virtual functions with energy higher than 200 a.u. at double-zeta basis set contributes around 1.67\% 
into HFS constant, 1.86\% to $E_\mathrm{eff}$, and 1.96\% to $W_s$. Therefore, the error caused by the exclusion of high-lying virtual
spinors with energy more than 200 a.u. in our results with basis D is expected to be within 2\%. 
The error caused by the neglect of the higher-order relativistic effects can be avoided by adding the higher-order relativistic terms (for example, Gaunt
interaction term) in the Hamiltonian. 
The mean field Gaunt correction to our results can be estimated from Table \ref{core_effect} by comparing the values obtained using the Dirac-Coulomb and
the Dirac-Coulomb-Gaunt Hamiltonian. The Gaunt term contributes  around 1.5\% to $E_\mathrm{eff}$ and 0.3\% to $W_s$ with DZ basis.
The error due to the missing higher-order correlation effects is reported as about 3.5\% for some other but similar system to BaF in 
literature \cite{kudashov_2014, das_hgf}. We, therefore, expect a similar amount of error due to missing higher-order correlation effects
in the BaF molecule. On the other hand, vibrational effects
are usually important for those molecular properties which strongly depend on the internuclear distance of the molecule. As the matrix elements of AIC properties are highly 
concentrated near the heavy nucleus, we do not expect a significantly large error due to the neglect of vibrational effects.
Furthermore, the error due to the basis set incompleteness can be estimated from Table \ref{baf_pt} by comparing the values obtained using
DZ basis with those obtained with TZ basis or TZ basis with QZ basis. The change of magnitude is about 3.64\% for $E_\mathrm{eff}$ and 9.35\% for $W_s$ when we improve
the quality of basis from DZ (i.e., A) to TZ (i.e., C). Similarly, for the improvement in basis from TZ (B) to QZ (D), the magnitude of $E_\mathrm{eff}$ and $W_s$ changes
by around 0.6\% and 0.6\%, respectively. Thus, we don't expect that the error due to basis set incompleteness would exceed 1\%. 
It is worth mentioning that although we correlate all the electrons to compute our most reliable results, they are not completely free from the error associated with core correlation effects since we use dyall.cv4z basis sets, which are designed only to treat core-valence correlation. However, relying on our previous studies \cite{sudip_pbf_pt, talukdar_hfs, talukdar_hgh} we expect this error to be negligible for BaF as well.
Nevertheless, ignoring
the mutual cancellations of different possible sources of error and assuming the errors 
to be independent, we can argue that the uncertainty in our most reliable results for the $\mathcal{P,T}$-odd molecular parameters of BaF is within 8\%.


\par



\section{Conclusion} \label{conc}
We report the $\mathcal{P,T}$-odd interaction coefficients and the hyperfine structure constants of BaF along with its
molecular frame dipole moment employing the four-component relativistic CCSD method. The good agreement of the calculated 
HFS constant with available experimental results reflects the accuracy of the wave function near the nuclear region generated by the employed method.
The most reliable values of $E_\mathrm{eff}$ and $W_s$ in BaF reported by us are 6.52 GV/cm and 8.35 kHz, respectively, within
an uncertainty of 8\%. 

\section*{Acknowledgement}
We acknowledge the resources of the Center of Excellence in Scientific Computing at CSIR-NCL. 
K.T. is grateful to Dr. Himadri Pathak and Dr. Sudip Sasmal for their insightful suggestions. 
K.T. thanks IIT Bombay for the Research Associateship.
\vspace{1cm}

\end{document}